# Electrical control of the valley Hall effect in bilayer MoS$_2$ transistors

Jieun Lee, Kin Fai Mak*, and Jie Shan*
Department of Physics and Center for 2-Dimensional and Layered Materials,
The Pennsylvania State University, University Park, Pennsylvania 16802-6300, USA
*E-mails: kzm11@psu.edu; jus59@psu.edu

**The valley degree of freedom of electrons in solids has been proposed as a new type of information carriers beyond the electronic charge and spin[1-6]. Recent experimental demonstrations of the optical orientation of the valley polarization[7-10] and generation of the valley current through the valley Hall effect[11] in monolayer MoS$_2$ have shown the potential of two-dimensional semiconductor transition metal dichalcogenides for valley based electronic and optoelectronic applications. The valley Hall conductivity in monolayer MoS$_2$, a non-centrosymmetric crystal, however, cannot be easily tuned, presenting a challenge for valley-based applications. Here we report the control of the valley Hall effect in bilayer MoS$_2$ transistors through a gate. The inversion symmetry present in bilayer MoS$_2$ was broken by the gate applied electric field perpendicular to the plane. The valley polarization near the edges of the device channels induced by the longitudinal electrical current was imaged by use of Kerr rotation microscopy. The polarization is out-of-plane, has opposite sign for the two edges, and is strongly dependent on the gate voltage. The observation is consistent with the symmetry dependent Berry curvature and valley Hall conductivity in bilayer MoS$_2$[12]. Our results are another step towards information processing based on the valley degree of freedom.**

Monolayers of group-VI transition metal dichalcogenides (TMDs) of the hexagonal polytype consist of a single layer of transition metal atoms sandwiched between two layers of chalcogen atoms in a trigonal prismatic structure (Fig. 1a). Because of the absence of inversion symmetry, electrons in monolayer TMDs at the K and the K' valley of the Brillouin zone possess finite but opposite Berry curvatures[3,13]. This feature of the electronic Bloch states gives rise to interesting topological transport properties such as the valley Hall effect (VHE), which has recently been observed in monolayer molybdenum disulfide (MoS$_2$)[11] as well as gapped graphene[14-16]. In the VHE (Fig. 1b), a valley current density (net current density of the K and K' electrons) $J_V = J_K - J_{K'}$ is induced in the transverse direction by a longitudinal electrical current under a source-drain bias field $E_x$. The valley Hall conductivity $\sigma_{VH}=J_V/E_x$ arisen from the Berry curvature effect, however, cannot be easily tuned in large gap materials such as monolayer MoS$_2$. On the other hand, in bilayer MoS$_2$, which is made of two equivalent monolayers rotated by $180^0$ with respect to each other, the inversion symmetry is restored (Fig. 1a). The Berry curvature effects vanish and no VHE is observed[11]. The inversion symmetry, however, can be broken by application of an electric field perpendicular to the plane. Finite Berry curvatures and valley Hall conductivity $\sigma_{VH}$ can be created. Furthermore, the sign of $\sigma_{VH}$ is expected to change upon reversing the perpendicular field direction. This phenomenon can be thought of as a complete or partial cancelation of the valley Hall currents of opposite direction in two constituent monolayers of the bilayer system. Bilayer MoS$_2$ thus presents an electrically highly tunable system for valley transport.

Since electrical methods do not directly probe the presence of a valley Hall current (equal flow of the K and K' electrons in opposite direction produces zero Hall voltage), we employ Kerr rotation (KR) microscopy to measure the valley polarization near the channel edges driven by the valley Hall current. The method has been used in similar studies of the spin Hall effect (SHE), for instance, in GaAs[17,18]. In this method, linearly polarized light is focused onto a sample under normal incidence. The polarization of the reflected beam is rotated by an amount that is proportional to the net magnetization of the sample in the out-of-plane direction (Fig. 1b). Instead of detecting the spin magnetization in the SHE, here we measure the valley (or orbital) magnetization. More specifically, under steady-state conditions, the valley Hall current density $J_V$ is balanced by intervalley relaxations at the sample boundaries, $J_V = en_V \lambda_V / \tau_V$, where $e$ is the electron charge, $n_V = n_K - n_{K'}$ is the intervalley (or K and K' electron) population density imbalance, $\lambda_V$ is the valley mean free path, and $\tau_V$ is the intervalley scattering time. As a result, net valley polarizations of opposite sign are accumulated at the two edges of the device channel. For bilayer MoS$_2$ under a vertical electric field, the two constituent monolayers are at different potentials (Fig. 1c). Since left (right) circularly polarized light only couples to electrons in the K (K') valley of the top (bottom) layer and the K' (K) valley of the bottom (top) layer[19], a finite KR angle due to the relative phase shift between left and right circularly polarized light is expected when $n_V \neq 0$.

Experiments were performed on several bilayer MoS$_2$ field-effect transistor (FET) devices that were fabricated by mechanical exfoliation of bulk MoS$_2$ crystals followed by direct transfer onto SiO$_2$/Si substrates with pre-patterned Ti/Au electrodes (Fig. 1d). This simple fabrication process avoids degradation of the material by lithographic processes and provides devices with good optical and electrical characteristics. Figure 1e shows the gate voltage $V_g$ dependence of the two-point conductivity $\sigma_{xx}$ of a typical bilayer device, where the Si substrate is used as a back gate. N-type conduction is seen in MoS$_2$, consistent with previous reports[20,21]. The bias dependences of the current density $J$ under different gate voltages (inset, figure 1e) are nonlinear, indicating the presence of Schottky barriers at the semiconductor-metal contacts. The longitudinal conductivity was estimated using $\sigma_{xx} \approx J/E_x$, where $E_x \approx V_x/l$ is approximated by the bias voltage $V_x$ from the linear regime and the channel length $l$. From the gate dependence of the conductivity we also estimated the electron field-effect mobility $\mu \approx 120$ cm$^2$V$^{-1}$S$^{-1}$ using $C_g \approx 3.6 \times 10^{-8}$ Fcm$^{-2}$ for the Si back gate capacitance. These estimates provide a lower bound for the conductivity and mobility. Below we measure the spatial dependence of the Kerr rotation angle $\delta\theta$ of bilayer MoS$_2$ FETs under different bias and gate voltages. Unless otherwise specified, all measurements were performed in high vacuum at 30 K. For more details on sample preparation and experimental setup as well as the optical properties of the devices, refer to Method and Supplementary Materials Section 1.

Figure 2b shows the two-dimensional (2D) map of $\delta\theta$ for bilayer device #1 under a gate voltage of $V_g = 20$ V. The KR angle was detected using a lock-in technique under a modulating bias voltage $V_x$ with a peak-to-peak value of 2.5 V. The channel boundaries are indicated with dashed lines. For comparison, the direct reflection image of the device is shown in figure 2a. Finite KR signal is clearly detectable only near the two edges of the channel and has opposite sign at the two edges. The spatial resolution of the images is ~ 0.7 μm and the KR signal is at the level of ~ 100 μrad along the two edges. We have performed multiple control experiments on multiple devices to verify the origin of the observed signal. These tests include the directional dependence of the KR signal on the bias field, a nil result for pure modulation of the gate voltage (in the absence of a bias field), the temperature dependence and the probe wavelength dependence. (See Supplementary Materials Section 2, 3 for details.) A systematic study of the KR angle on the bias voltage was then performed near one of the edges (left axis, Fig. 2e). Also shown is the $J-V_x$ dependence under the same gate voltage (right axis, Fig. 2e). The two quantities are seen to follow the same dependence on $V_x$, indicating that the observed KR is driven by and linearly proportional to the longitudinal current.

Next we investigate KR imaging of the device under different gate voltages. Results for two voltages ($V_g$ = -5 and 4 V) are shown in Fig. 2c and d. At $V_g$ = -5 V, for which the vertical electric field reverses its direction, the KR is again detectable only near the two channel edges, but switches sign from that under $V_g$= 20 V. At the intermediate voltage of $V_g$ = 4 V, however, besides small spatial inhomogeneities, no obvious KR is detected near the edges of the device channel. The $V_g$-dependence of the KR signal is more systematically studied in the contour plot of figure 3a for a single horizontal line containing both channel edges under a gate voltage varying in small steps. The result further confirms that if small spatial inhomogeneities throughout the device are ignored, the KR signal is present only near the channel edges, has opposite sign for the two edges, and depends strongly on the gate voltage. The latter is highlighted in the gate dependence of the KR at the two edges in Fig. 3b. The KR angle starts from a small value at large negative $V_g$, increases in magnitude, goes through a peak, switches sign and continues to increase in magnitude while $V_g$ is varied from -20 V to 20 V.

The observed dependences of the KR on the bias voltage, gate voltage and spatial coordinates are fully compatible with the presence of a gate tunable VHE in bilayer $MoS_2$ FETs (Fig. 2f). We note that although the observed out-of-plane magnetization could also be generated by the SHE, such an effect is negligible in *n*-doped $MoS_2$ due to the small spin-orbit coupling in the conduction bands near the K and K' point of the Brillouin zone[5,22]. For inversion symmetric bilayers, the Berry curvature is zero, no VHE is present and the KR is zero. When a vertical electric field breaks the inversion symmetry, the Berry curvature (which is an odd function of the vertical field) is no longer zero[12], the VHE is present and the KR is finite. When the vertical electric field changes sign, the Berry curvature, the valley Hall conductivity $\sigma_{VH}$, as well as the valley polarization accumulated at the channel edges all change sign. For device #1, $V_g$ = 4 V corresponds to the gate voltage under which the change of sign occurs, suggesting that 4 V is required to bring the bilayer sample on Si substrate back to inversion symmetry. Such an interpretation is reasonable if one considers spontaneous doping of the bilayer from the Si substrate, a phenomenon that has been widely recognized for atomically thin samples on Si substrate[23].

To independently verify the above interpretation, we have performed second harmonic generation (SHG) studies on our bilayer $MoS_2$ devices. The SHG process described by a third-rank tensor $\chi^{(2)}$ in the electric dipole approximation is a sensitive probe of the sample symmetry[24-26]. For inversion symmetric bilayer $MoS_2$ (belonging to the $D_{3d}$ point group), all $\chi^{(2)}$ elements are zero and the SHG vanishes. However, under a vertical electric field, the inversion symmetry of the bilayer (now belonging to the $C_{3v}$ point group) is broken, non-vanishing $\chi^{(2)}$ elements emerge and the SHG is generally nonzero. To enhance the detection sensitivity for excitation under normal incidence, we have employed vertically polarized light to excite the sample and collected the reflected second harmonic without resolving its polarization. This was achieved by using a radial polarization converter in combination with an objective lens[27,28] (See Method for more details). Figure 4b shows the spectrally integrated SHG intensity as a function of $V_g$ for device #2. The SHG minimum occurs at $V_g \approx 15$ V and around it, the SHG varies quadratically with $V_g$. Also shown in Fig. 4a is the $V_g$-dependence of the KR signal on one of the channel edges of the same device under different bias voltages. It is clear that the point at which the KR signal changes sign coincides with the point at which the SHG is minimum. This result verifies that the observed valley polarization in bilayer MoS$_2$ is correlated with inversion symmetry breaking, a necessary requirement for the observation of the VHE.

Finally, for quantitative understanding of our experimental results we compare experiment with the predictions of a simple $\boldsymbol{k} \cdot \boldsymbol{p}$ model[5,12] that includes the two lowest energy bands for the field dependent Berry curvature effects in bilayer MoS$_2$ (See Supplementary Materials Section 4 for detailed derivations). KR can be related with the complex absorbance difference between left and right circularly polarized light $\alpha_+ - \alpha_-$ for bilayer MoS$_2$: $\delta\theta = Im[\beta(\alpha_+ - \alpha_-)]$ by performing a wave propagation analysis in the multilayer system consisting of bilayer MoS$_2$, SiO$_2$ thermal layer and Si substrate[29]. Here $\beta$ is a parameter that is determined by the layer thicknesses and local field factors at the probe wavelength. The absorbance difference originated from a small intervalley population imbalance $n_V$ at the channel edges can be expressed as $\alpha_+ - \alpha_- = \frac{\lambda_V}{d}\frac{d\alpha}{dn}n_V$. Here the ratio between the valley mean free path $\lambda_V$ and the size of the probe beam $d$ takes into account the fact that $n_V$ only spans a distance of $\lambda_V$ from the channel edges; the second factor $d\alpha/dn$, which accounts for the rate at which the optical absorbance of MoS$_2$ changes with doping, depends on $V_g$ weakly for the relatively small range of $V_g$ studied here[30]. We further invoke the definition of the valley Hall conductivity and express the KR in terms of $\sigma_{VH}$

$$\delta\theta = Im[\beta \frac{d\alpha}{dn}]\frac{\tau_V}{ed}\sigma_{VH}E_x. \tag{1}$$

For relatively small gate voltages, the $\boldsymbol{k} \cdot \boldsymbol{p}$ model predicts the following simple expression for the valley Hall conductivity: $\sigma_{VH} \propto n\frac{V}{\sqrt{t_{\perp c}^2+V^2}}$[12]. Here $t_{\perp c}$ is the interlayer hopping constant of the conduction band, and $V$ and $n$ denote the interlayer potential difference and the total doping density of the bilayer, respectively. The latter two parameters can be related to the gate voltage by solving the electrostatics problem of gating: $n = C_g(V_g - V_0)$ and $V \approx (C_g/C_Q)(V_g - V_s)$. Here $C_Q = \frac{2me^2}{\pi\hbar^2} \approx 10^{-4} Fcm^{-2}$ is the quantum capacitance of a single MoS$_2$ layer with electron band mass $m$, $V_0$ is the gate voltage at which the bilayer is charge neutral, and $V_s$ is the gate voltage at which the bilayer on substrate is inversion symmetric. The value of $V_s$ has been determined by two independent measurements above (KR and SHG). In Fig. 3b we show that the main gate dependence can be captured by ignoring all $V_g$-dependences in Eq. (1) except that of $n$ and $V$. The fit (solid line, Fig. 3b) including three free parameters, a scaling factor for $\delta\theta$, $V_0$, and $t_{\perp c}$ yields $t_{\perp c} \approx 2$ meV. Such a relatively small interlayer hopping constant is compatible with the symmetry of $d_{z^2}$ orbitals (predicting $t_{\perp c} = 0$) from which the conduction band at the K and K' point is mainly made of[19]. The nonzero interlayer hopping is likely due to the contribution of the sulphur *p*-orbitals. We note that although a more thorough analysis of the gate dependence of the absorbance, intervalley scattering time and electron mobility is required to fully describe the observed Kerr rotation, our experiment clearly demonstrates electrical control of the VHE in bilayer $MoS_2$. These results and the sensitive Kerr rotation microscopy developed here will open up new possibilities for control of topological transport in 2D materials and its studies.

**Methods**

Atomically thin flakes of $MoS_2$ were fabricated by mechanical exfoliation of bulk synthetic $MoS_2$ crystals (2D Semiconductors) onto polydimethylsiloxane (PDMS) substrate. The sample thickness was first calibrated by optical reflection contrast and then confirmed by photoluminescence and Raman spectroscopy[31]. Bilayer flakes of rectangular shape were transferred onto Si substrate with pre-patterned Ti/Au electrodes to form field-effect transistors (Fig. 1d). The flakes were aligned such that their long edges are perpendicular to the source and drain electrodes. Lightly doped Si substrate with a 100 nm thermal oxide layer was used as the back gate. All electrical and optical measurements were performed under high vacuum at 30 K.

Kerr rotation (KR) microscopy (figure 1b): Linearly polarized continuous wave radiation was focused onto $MoS_2$ devices using a 40x Olympus objective under normal incidence. The photon wavelength (654 nm) was slightly detuned to the red side from the A exciton resonance of bilayer $MoS_2$. The beam diameter on the sample is ~ 0.7 $\mu m$ and the power is no more than 250 μW. The reflected beam was collected by the same objective, passed through a half waveplate, a Wollaston prism, and detected by a pair of balanced photodiodes. The KR angle was determined as the ratio of the intensity imbalance of the photodiodes and the intensity of each photodiode. To enhance measurement sensitivity, lock-in detection of the KR signal was employed by modulating the bias voltage $V_x$ at 4.25 kHz. Relatively low bias voltages (with a peak-to-peak amplitude ≤ 2.5 V) were used to avoid any significant source-gate coupling effects. Two-dimensional imaging of the KR signal was achieved by scanning the device on a piezo XY scanner.

Second-harmonic generation (SHG): A femtosecond Ti:Sapphire oscillator that delivers optical pulses centered at 810 nm with a pulse duration of ~ 100 fs and a repetition rate of 80 MHz was employed. A linearly polarized beam was focused onto the $MoS_2$ devices by a 40x Olympus objective under normal incidence after passing through an s-waveplate. At the focal point most light is vertically (or z-) polarized and no more than 1.8 mW was used on the sample. The reflected beam was collected by the same objective, passed through a filter to remove the radiation at the fundamental frequency, and detected by a grating spectrometer equipped with a nitrogen cooled charge-coupled detector (CCD). Several spectra of the second harmonic (SH) radiation generated from device #2 are shown in the inset of Fig. 4b for different gate voltages. The integration time for each spectrum was 10 s. In this configuration, the main contribution to the SHG arises from the second-order nonlinear susceptibility tensor element $\chi^{(2)}_{zzz}$, which can also be expressed as the product of the third-order nonlinear susceptibility element $\chi^{(3)}_{zzzz}$ and the vertical electric field $E_z$ applied to the bilayer[32]. For small $V_g - V_s$, the effect due to change of doping is relatively small and the SH intensity scales quadratically with the field $\propto |\chi^{(3)}_{zzzz} E_z|^2 \propto (V_g - V_s)^2$ as shown by the solid line in Fig. 4b.

**Figures**

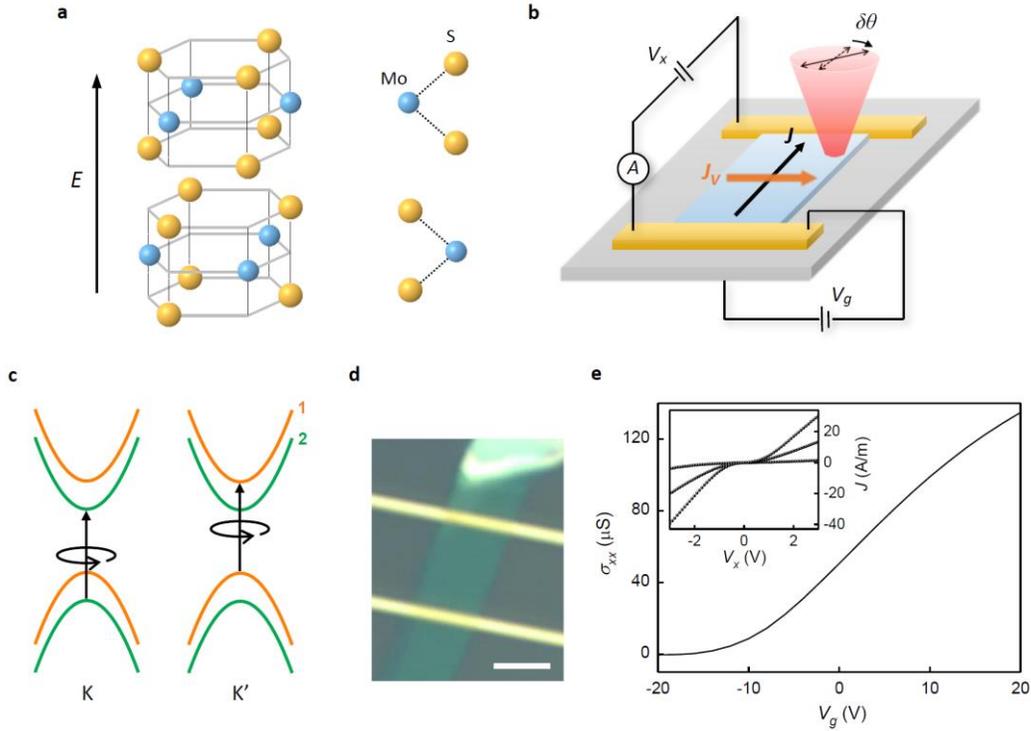

**Figure 1 | MoS$_2$ structure and device characterization. a**, Bilayer MoS$_2$ crystal structure and vertical electric field, *E*, that breaks the inversion symmetry of the bilayer. **b**, Schematics of a MoS$_2$ field-effect transistor with bias voltage ($V_x$) applied on the source-drain electrodes and gate voltage ($V_g$) applied through the Si/SiO$_2$ substrate. In the valley Hall effect, a valley current $J_V$ (orange arrow) in the transverse direction is induced by a longitudinal electrical current *J* (black arrow). The VHE is detected by focusing a linearly polarized probe beam onto the device under normal incidence and measuring the Kerr rotation angle δθ of the reflected beam. **c**, Band diagram of bilayer MoS$_2$ under a finite vertical field *E* (orange and green lines for electronic states associated with layer 1 and 2, respectively) and the optical selection rules for right-handed light near the K and K' point of the Brillouin zone. **d**, Optical image of device #1. Scale bar is 5 μm. **e**, Conductivity of device #1 as a function of $V_g$. Inset shows $J$–$V_x$ dependences for $V_g$ at -10, 0 and 10 V.

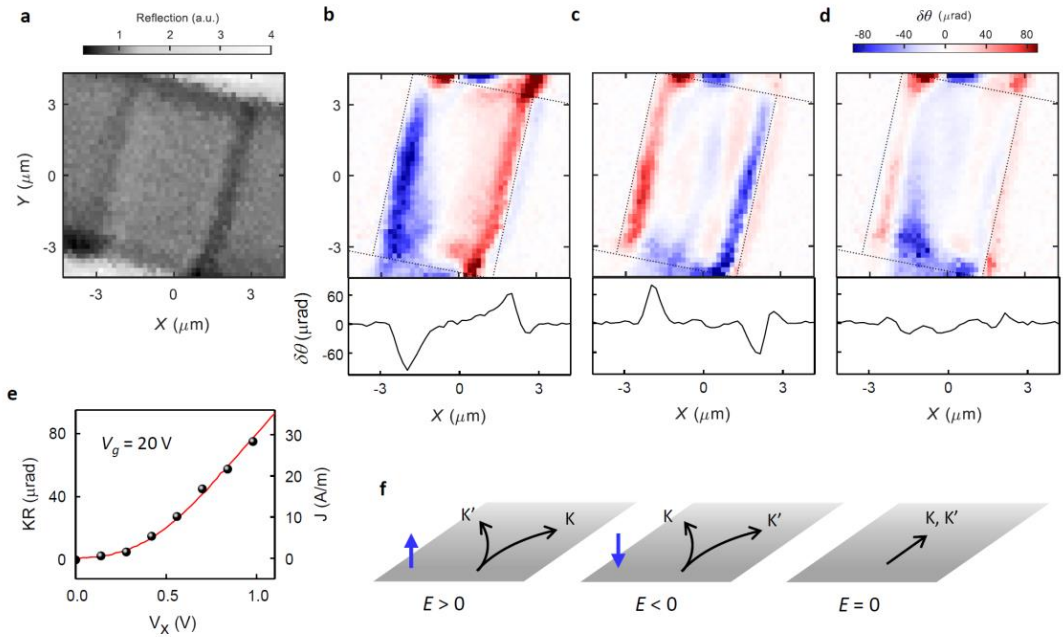

**Figure 2 | Valley Hall effect in bilayer MoS$_2$ (device #1). a**, Reflection image of the device. The signal is normalized to the reflectivity of the MoS$_2$ channel. **b-d**, Spatial 2D map and selected horizontal linecut of the Kerr rotation angle δθ under $V_g$ = 20 V (**b**), -5 V(**c**) and 4 V (**d**). All measurements were done under the same $V_x$. Black dashed lines show the boundaries of the channel determined from the reflection image. **e**, Bias voltage ($V_x$) dependence of δθ on one edge of the channel (black symbols, left axis) and the longitudinal current density $J$ (red line, right axis) under $V_g$ = 20 V. **f**, Illustration of the field-controlled valley Hall effect in bilayer MoS$_2$. Blue arrows show the direction of the vertical *E*-field.

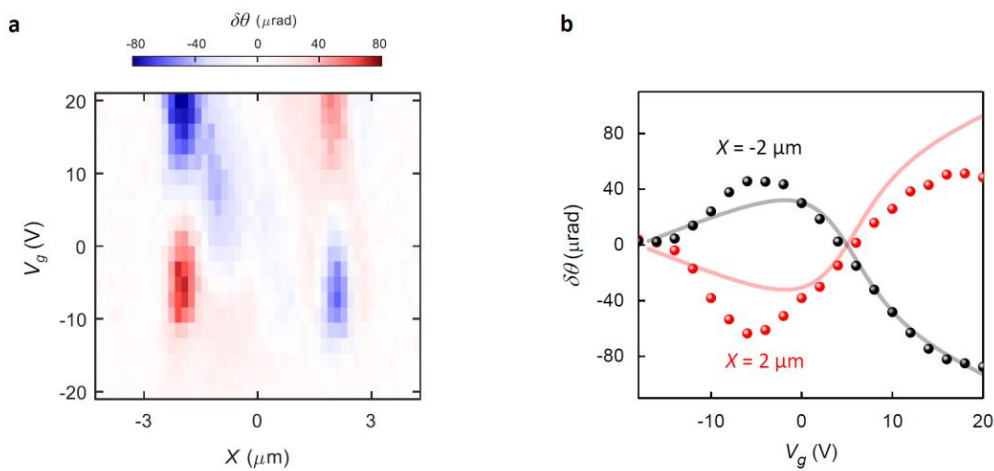

**Figure 3 | Gate dependence of the valley Hall effect (device #1). a**, Kerr rotation δθ map as a function of *X* position and gate voltage $V_g$. **b**, Gate dependence of δθ on the left

edge ($X = -2$ μm, black symbols) and right edge ($X = 2$ μm, red symbols). Solid lines are fit to the gate dependent VHE model as described in the text.

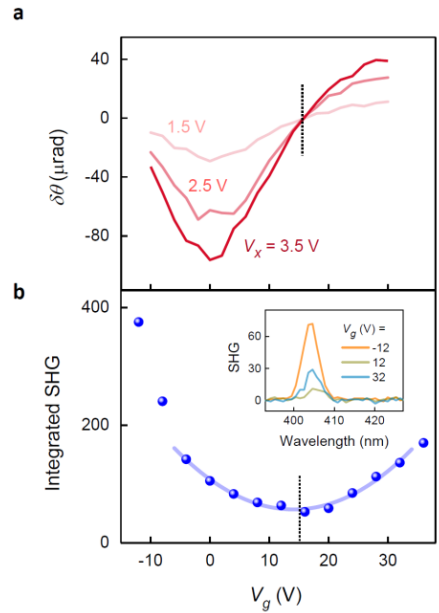

**Figure 4 | Gate dependence of the valley Hall effect and second harmonic generation (device #2). a**, Kerr rotation δθ measured on one edge of the device channel as a function of gate voltage at selected bias voltages. The black dashed line at $V_g \approx 15$ V indicates the gate voltage at which δθ changes sign. **b**, Gate dependence of the spectrally integrated SHG (symbols) and fit to a quadratic dependence (solid blue line). Minimum occurs at $V_g \approx 15$ V (dashed black line) where inversion symmetry of the bilayer is restored. Inset shows the spectra of SHG recorded at selected gate voltages.